\documentclass[12pt]{article}

\def\lromn#1{\uppercase\expandafter{\romannumeral#1}}
\def\ds{\displaystyle}
\def\vev#1{\langle 0 |#1|0 \rangle}

\usepackage{graphicx}

\begin{document}

\begin{flushright}
ICRR-Report-495-2002-13\\
YITP-02-68\\
  \today\\
\end{flushright}

\begin{center}

  \begin{large}
    \textbf{Unitarity and Higher-Order Corrections \\in Neutralino 
Dark Matter Annihilation \\ into Two Photons}
  \end{large}

  \vspace{2cm}

  \begin{large}
    Junji Hisano$~^a$, Sh. Matsumoto$~^a$, and Mihoko M. Nojiri$~^b$

    $~^a$
    ICRR,
    University of Tokyo,
    Kashiwa 277-8582,
    Japan
    \\
    $~^b$
    YITP, 
    Kyoto University, 
    Kyoto 606-8502, 
    Japan
  \end{large}

  \vspace{3cm}

  {\bf ABSTRACT}

\end{center}

The neutralino pair annihilation into two photons in our galactic halo
gives a robust dark matter signal, since it would give a
quasi-monotonic gamma ray.  This process is radiatively-induced, and
the full-one loop calculation was done previously.  However, for the
heavy wino-like or Higgsino-like neutralino, the one-loop cross
section violates unitarity, therefore the higher-order corrections may 
be important. We construct a non-relativistic theory for chargino and
neutralino two-body states, and estimate all-order QED corrections and
two-loop corrections by $Z$ and/or $W$ exchange. We find that the
critical mass, above that the two-loop contribution is larger than
one-loop one, is about 8 TeV ($O(10)$ TeV) in the limit where neutralino is
wino (Higgsino)-like, respectively.  Around and above the critical
mass, the all-order Z and/or W exchange must be included to estimate
the cross section. On the other hand, the QED corrections depend on
the mass difference between the neutralino and chargino. In the
wino-like limit where neutralino is highly degenerate with chargino in
mass, we find that QED corrections enhance the pair annihilation cross
section by 1.5-2.

\newpage


\lromn 1 \hspace{0.2cm} {\bf Introduction}

\vspace{0.5cm}

Present observation of cosmological and astrophysical quantities
allows a precise determination of the mean density of matter
($\Omega_M$) and baryon density ($\Omega_B$) in the Universe, and the
existence of non-baryonic dark matter (DM) is established now
\cite{Turner:2001mw}.  However, the constituent of the DM is still
an unresolved problem. The supersymmetric (SUSY) models provides the
candidates of the DM, since the lightest SUSY particle (LSP) is stable
due to the conserved R parity. The LSP may be the lightest neutralino
in minimal supersymmetric standard model (MSSM). It is a linear
combination of gauginos (bino and wino) and Higgsinos, which are
superpartners of gauge and Higgs bosons, respectively. Thermal
\cite{Ellis:1983ew} or
non-thermal processes \cite{non-thermal} in the early Universe may
produce the lightest neutralino enough to explain the DM in the
Universe.

The detection of exotic cosmic rays is feasible technique to search
for the dark matter particles, since a pair of the lightest neutralino
could annihilate into the SM particles with significant cross section
\cite{Silk:1985ax}.  Among those, excess of monochromatic gamma ray
due to the neutralino annihilation into two photons is a robust
signal if observed, because the diffused gamma-ray background must
have a continuous energy spectrum \cite{Bergstrom:1997fj}.

The neutralino annihilation to two photons is a radiative process.
The dominant contribution to the cross section comes from the process
where a neutralino pair is converted into a virtual chargino pair by 
$W$-boson exchange and then the chargino pair annihilates into two
photons.  The full one-loop cross section is calculated in
Ref.~\cite{Bergstrom:1997fh}.  The surprising fact is that the cross
section is suppressed only by the $W$-boson mass ($m_W$), not by the
neutralino mass ($m$), as $ \sigma v
\sim {\alpha^2\alpha_2^2}/{m_W^2}$, if the lightest-neutralino
mass eigenstate is very close to wino or Higgsino. Since other
pair-annihilation cross sections are proportional to $1/m^2$, the
potential of search for the monochromatic gamma ray in cases of the wino-
or Higgsino-like neutralino DM then can be understood as follows; as
the neutralino mass increases, the signal rate of the monochromatic
gamma ray reduces less quickly compared with the other DM signatures.

On the other hand, this cross section should be bounded from above by
the unitarity limit $\sigma v < 4\pi/(v m^2)$ \cite{Jacob:at}. Thus,
for the extreme heavy neutralino, this one-loop result must fail
\cite{Drees:1998ra}, and the higher-order corrections should be
included.  Indeed, the wino- or Higgsino-like neutralino is
accompanied with  chargino in the same  SU(2) multiplet, and their masses are
almost degenerate. In these cases, the intermediate chargino-pair
state in the process is almost on-shell, since the neutralino DM is
non-relativistic (NR). When $m$ is large, the diagrams are enhanced by
a factor of $\alpha_2 m/m_W$ for each $W$ boson exchange, and the
higher-order loop diagrams become more and more important. Then, the
one-loop result is not valid anymore, and we need to sum the
contributions from ladder diagrams of the weak-boson exchange to all
orders.

This failure of the perturbative expansion is similar to the threshold
singularity in pair creation or annihilation processes in QCD or QED
\cite{Appelquist:zd}.  In the quark-pair annihilation process at the
threshold (NR) region, the higher-order corrections are non-negligible
since nearly on-shell quark is dominated in the loop integration due
to the small relative velocity. In other words, the deformation of the
quark wave function by the QCD potential is not negligible at the
threshold region. In order to get the reliable cross section, we need
to sum ladder diagrams by gluon exchange to all orders, or to use
the wave function for quarks under the QCD potential.  In the
neutralino annihilation to two photons, we need to include the Yukawa
potential of the weak boson. When $1/(\alpha_2 m)$ is larger than the
effective range of the Yukawa potential $1/m_W$, the Yukawa potential
is point-like in the coordinate space, and does not deform the wave
functions of neutralino and chargino. However, if $m$ is heavier, the
wave functions of chargino and neutralino are deviated from plane
waves inside $\sim 1/(\alpha_2 m)$.  In this case, the perturbative
expansion at the threshold region should be broken.

The NR field theory is useful to investigate these higher-order
corrections at the threshold region in QCD or QED
\cite{Caswell:1985ui}. In this technique, we can factorize the
short-distance physics, such as annihilation or production, from the
long-distance physics related to the wave function
\cite{Bodwin:1994jh}. The evaluation of the contribution from the
long-distance physics is possible either by systematical resummation
of the ladder diagrams in diagrammatic methods or by evaluating
the wave function under the potential.  This formalism is also useful
to evaluate the higher-order corrections in the neutralino annihilation
to two photons.  From a viewpoint of the NR field theory, the
short-distance physics comes from the chargino-pair annihilation to
two photons, and the long-distance physics from the weak-boson
exchanges, in addition to the photon exchange between charginos.

In this paper, the higher-order corrections to the annihilation cross
section of the neutralino to two photons is studied in cases of the
wino- or Higgsino-like neutralino DM. First, we construct the two-body
effective action of non-relativistic neutralino and chargino pairs.
Using this formula, we estimate two-loop corrections by  weak-boson
exchange in a diagrammatic method.  The two-loop contribution becomes
important when the neutralino is as heavy as $m_{W}/\alpha_2$. From
the explicit calculation, we found that the critical mass, above that
the two-loop contribution is larger than one-loop one, is about 8 TeV
for the wino-like neutralino, and is about $O(10)$ TeV for the Higgsino-like
one.  The resummation of the weak-boson exchange contributions to all
orders is possible by solving numerically the wave functions of
chargino and neutralino under the Yukawa potentials of the weak bosons
\cite{work_in_progress}. We also calculate all-order corrections by
photon exchange, by using the wave function deformed by Coulomb
potential. The QED corrections depend on the mass difference between
neutralino and chargino, $\delta m$. The wino-like neutralino
is highly degenerate with the chargino, and $\delta m$ is typically
$\sim$ 0.1 GeV. For this case, we found that the QED corrections
enhance the cross section by 1.5$\sim$2 when the neutralino mass is a
few TeV.
 
This paper is organized as follows. In next section we briefly review
masses and interactions of neutralino and chargino, especially
in the wino- and Higgsino-like cases.  In section III, we construct
the two-body effective action of the NR neutralino and chargino pairs
and show the strategy for our calculation. Here, in order to show the
validity for our formalism, we reproduce the previous one-loop
result, and show that the unitarity bound is satisfied in an extremely
heavy neutralino mass limit. In section IV, we first include all-order
QED corrections to the leading cross section, and we evaluate two-loop
corrections by $Z$ and $W$ boson exchange. Section V is devoted to
conclusion and discussion.  We give a full effective Lagrangian
relevant to our study in Appendix A.
\vspace{1cm}


\lromn 2 \hspace{0.2cm} {\bf Non-Relativistic  Action of 
Wino- or Higgsino-like Neutralino and Chargino.}

\vspace{0.5cm} 

In this section, we review the mass spectrum and the low-energy
interaction of the SU(2) multiplets containing the lightest
neutralino. We are interested in the case where the lightest
neutralino LSP ($\tilde{\chi}^0_1$) is almost degenerate with the
lighter chargino ($\tilde{\chi}^+\equiv \tilde{\chi}^+_1$) in mass.
For simplicity, we assume that all the other SUSY particles, including
the two heavier neutralino states ($\tilde{\chi}^0_3$ and
$\tilde{\chi}^0_4$) and the heavier chargino ($\tilde{\chi}^+_2$), are
so heavy that we can ignore their contributions to the
DM annihilation.

Neutralinos are linear combinations of supersymmetric partners of
gauge bosons, the bino ($\tilde{B}$) and the neutral wino
($\tilde{W}^0$), and those of Higgs bosons, the Higgsinos
($\tilde{H}_1^0$, $\tilde{H}_2^0$). While those four fields have
SU(2)$\times$U(1) invariant masses, they are mixed with each others by
the SU(2)$\times$U(1) symmetry breaking. The lightest neutralino is
wino-like when $M_2\ll \mu, M_1$ and Higgsino-like if $\mu\ll M_1,
M_2$. Here $M_{1}$ and $M_{2}$ are the bino and the wino masses,
respectively, and $\mu$ is the supersymmetric Higgsino mass parameter.
When $M_1,M_2,\mu \gg m_Z$, the lightest neutralino mass is
\begin{eqnarray}
  m_{\tilde{\chi}^0_1}
  \simeq
  M_2 + \frac{m_Z^2 c_W^2}{M_2^2 - \mu^2} (M_2 + \mu \sin{2\beta})
\end{eqnarray}
for the wino-like case, and 
\begin{eqnarray}
  m_{\tilde{\chi}^0_1}
  \simeq
  \mu + \frac{m_Z^2(1 + \sin{2\beta})}
             {2(\mu - M_1)(\mu - M_2)}(\mu - M_1 c_W^2 - M_2 s_W^2)
\end{eqnarray}
for the Higgsino-like case ($\mu>0$). Here, we show the terms up to
${\cal O}({m_Z^2}/{m_{SUSY}})$.

Charginos are linear combinations of the charged wino, 
$\tilde{W}^{\pm}$, and the charged Higgsino ($\tilde{H}_1^-$,
$\tilde{H}_2^+$).
When $M_2,\mu \gg m_W$,
the lighter chargino mass is 
\begin{eqnarray}
  m_{\tilde{\chi}^+}
  \simeq
  M_2 + \frac{m_W^2}{M_2^2 - \mu^2} (M_2 + \mu \sin{2\beta})
\end{eqnarray}
for the wino-like case, and 
\begin{eqnarray}
  m_{\tilde{\chi}^+}
  \simeq
  \mu - \frac{m_W^2}
             {M_2^2-\mu^2}(\mu + M_2 \sin{2\beta})
\end{eqnarray}
for the Higgsino-like case.  We show the terms up to ${\cal
O}({m_Z^2}/{m_{SUSY}})$ here, again.

The mass difference between chargino and the LSP, $\delta
m$, is an important parameter for the neutralino annihilation cross
section into two photons.  For the wino-like case, their masses are
highly degenerate. The tree-level mass difference is ${\cal
O}(m_Z^4/m_{SUSY}^3)$ in a case $m_Z, M_2\ll M_1,\mu$, 
\begin{eqnarray}
  \delta m_{\rm tree}
  \simeq
  \frac{m_Z^4}{M_1\mu^2}s_W^2 c_W^2 \sin^22\beta~,
\end{eqnarray}
since the ${\cal O}(m_Z^2/m_{SUSY})$ corrections to the masses are
SU(2)$\times$U(1) invariant. The mass splitting receives the positive
contribution from the gauge boson loops due to the custodial SU(2)
breaking.  The radiative mass difference in the wino limit is
\begin{eqnarray}
  \delta m_{{\rm rad}}
  =
  \frac{\alpha_2 M_2}{4\pi} 
  \left(f(\frac{m_W}{M_2})-c_W^2f(\frac{m_Z}{M_2})-f(0)\right)~,
\end{eqnarray}
where $f(a) =\int^1_0 dx~ 2(1+x)\log(x^2+(1-x) a^2)$ 
\cite{Cheng:1998hc}. This correction is about $0.18$GeV when $M_2\gg m_W$.

For the Higgsino-like LSP, the mass splitting is 
${\cal O}(m_Z^2/m_{SUSY})$,
\begin{eqnarray}
  \delta m \simeq 
  \frac{1}{2} \frac{m_Z^2}{M_2}c_W^2 (1 - \sin2\beta)
  +
  \frac{1}{2} \frac{m_Z^2}{M_1}s_W^2 (1 + \sin2\beta)~.
\end{eqnarray}
The second-lightest neutralino ($\tilde{\chi}^0_2$) also degenerates
with the LSP and chargino, since they are in common SU(2) multiplets.
The mass difference between the LSP and the second lightest
neutralino, $\delta m_N$, is again ${\cal O}(m_W^2/m_{SUSY})$,
\begin{eqnarray}
  \delta m_N
  =
  \frac{m_Z^2}{M_2}c_W^2 + \frac{m_Z^2}{M_1}s_W^2~.
\end{eqnarray}
This is roughly $2\times \delta m$ when $\tan\beta\gg 1$. 	

Next, we consider interactions among neutralino(s) and chargino.
After ignoring all SUSY particles except $\tilde{\chi}^0_{1(2)}$ and
$\tilde{\chi}^+$, the interactions which we have to take into account
are only the gauge interactions. The leading correction by
non-vanishing $\delta m$ to the neutralino annihilation cross section
to two photons is ${\cal O}(\sqrt{\delta m})$. As will be shown in
next section, the ${\cal O}(\sqrt{\delta m})$ correction originates
from an infrared behavior of the loop integrand which is controlled by
$\delta m$.  The next-to-leading order correction is ${\cal O}({\delta
m})$. To this order, the Yukawa interactions of the Higgs bosons or
unphysical (NG) bosons must be included to keep the gauge
invariance. In this paper, we calculate the annihilation cross section
up to the leading-order correction ${\cal O} (\sqrt{\delta m})$,
however, it is straight-forward to include the next-to-leading order
calculations.

In the wino limit, the gauge interactions of inos
are 
\begin{eqnarray}
  {\cal L}_{\rm int}
  &=&
  -\frac{e}{s_W}
  \left(
    \overline{\tilde{\chi}^0_1}\gamma^\mu\tilde{\chi}^-W^\dagger_\mu
    +
    h.c.
  \right)
  +
  e \frac{c_W}{s_W}\overline{\tilde{\chi}^-}\gamma^\mu\tilde{\chi}^-Z_\mu
  +
  e\overline{\tilde{\chi}^-}\gamma^\mu\tilde{\chi}^-A_\mu~.
\end{eqnarray}
The NR Lagrangian can be derived by taking a NR limit of neutralino
and chargino and integrating out of the gauge fields, and we find
\begin{eqnarray}
  {\cal L}_{\rm NR}^{(W)}
  &=&
  \eta_N^\dagger
  \left(
    i\partial_t + \frac{\nabla^2}{2m}
  \right)\eta_N
  +
  \eta_C^\dagger
  \left(
    i\partial_t - \delta m + \frac{\nabla^2}{2m}
  \right)\eta_C
  +
  \xi_C^\dagger
  \left(
    i\partial_t - \delta m - \frac{\nabla^2}{2m}
  \right)\xi_C
  \nonumber \\
  &+&
  \frac{\alpha}{2}\int d^3y~
  \eta^\dagger_C(x)\xi_C(\vec{y},t)
  \frac{1 - \frac{c_W^2}{s_W^2}e^{-m_Z|\vec{x} - \vec{y}|}}
         {|\vec{x} - \vec{y}|}
  \xi_C^\dagger(\vec{y},t)\eta_C(x)
  \nonumber \\
  &-&
  \frac{\alpha}{2s_W^2}\int d^3y~
  \left(
    \eta_N^\dagger(x)\eta_N^c (\vec{y},t)
    \frac{e^{-m_W|\vec{x} - \vec{y}|}}
         {|\vec{x} - \vec{y}|}
    \xi_C^\dagger(\vec{y},t)\eta_C(x)
    + h.c.
  \right)~.
  \label{winoNRL}
\end{eqnarray}
Here, $\eta_C$, $\xi_C$ and $\eta_N$ are defined as
\begin{eqnarray}
  \eta_C = \frac{1 + \gamma^0}{2}\tilde{\chi}^-e^{imt}~,
  \qquad
  \xi_C = \frac{1 - \gamma^0}{2}\tilde{\chi}^-e^{-imt}~,
  \qquad
  \eta_N = \frac{1 + \gamma^0}{2}\tilde{\chi}^0_1e^{imt}~.
\end{eqnarray}
We keep isospin singlet terms only, because only these terms are
necessary to calculate the neutralino annihilation cross section.

The gauge interactions in the Higgsino limit are
\begin{eqnarray}
  {\cal L}_{\rm int}
  &=&
  -\frac{e}{2s_W}
  \left(
    \overline{\tilde{\chi}^0_1}\gamma^\mu\tilde{\chi}^-W^\dagger_\mu
    -\overline{\tilde{\chi}^0_2}\gamma^\mu \gamma_5 \tilde{\chi}^-W^\dagger_\mu
    +
    h.c.
  \right)
  -
  \frac{e}{s_Wc_W} \left(\frac{1}{2} - c_W^2\right)
  \overline{\tilde{\chi}^-}\gamma^\mu\tilde{\chi}^-Z_\mu
  \nonumber \\
  &~&+~e
  \overline{\tilde{\chi}^-}\gamma^\mu\tilde{\chi}^-A_\mu
  +
  \frac{e}{2s_Wc_W}
  \overline{\tilde{\chi}^0_1}\gamma^\mu\gamma_5\tilde{\chi}^0_2Z_\mu~.
\end{eqnarray}
The NR Lagrangian is derived in the same way as the wino-like
case,
\begin{eqnarray}
  {\cal L}_{\rm NR}^{(H)}
  &=&
  \eta_N^\dagger
  \left(
    i\partial_t + \frac{\nabla^2}{2m}
  \right)\eta_N
  +
  \eta_C^\dagger
  \left(
    i\partial_t - \delta m + \frac{\nabla^2}{2m}
  \right)\eta_C
  \nonumber \\
  &+&
  \xi_C^\dagger
  \left(
    i\partial_t - \delta m - \frac{\nabla^2}{2m}
  \right)\xi_C
  +
  \xi_N^\dagger
  \left(
    i\partial_t - \delta m_N + \frac{\nabla^2}{2m}
  \right)\xi_N
  \nonumber \\
  &+&
  \frac{\alpha}{2}\int d^3y~
  \eta^\dagger_C(x)\xi_C(\vec{y},t)
  \frac{1 - \frac{(1 - 2c_W^2)^2}{4c_W^2s_W^2}e^{-m_Z|\vec{x} - \vec{y}|}}
         {|\vec{x} - \vec{y}|}
  \xi_C^\dagger(\vec{y},t)\eta_C(x)
  \nonumber \\
  &-&
  \frac{\alpha}{8s_W^2}\int d^3y~
  \left(
    \eta_N^\dagger(x)\eta_N^c (\vec{y},t)
    \frac{e^{-m_W|\vec{x} - \vec{y}|}}
         {|\vec{x} - \vec{y}|}
    \xi_C^\dagger(\vec{y},t)\eta_C(x)
    + h.c.
  \right)
  \nonumber \\
  &-&
  \frac{\alpha}{8s_W^2}\int d^3y~
  \left(
    \xi_N^\dagger(x)\xi_N^c (\vec{y},t)
    \frac{e^{-m_W|\vec{x} - \vec{y}|}}
         {|\vec{x} - \vec{y}|}
    \xi_C^\dagger(\vec{y},t)\eta_C(x)
    + h.c.
  \right)
  \nonumber \\
  &-&
  \frac{\alpha}{16c_W^2s_W^2}\int d^3y~
  \left(
    \eta_N^\dagger(x)\eta_N^c (\vec{y},t)
    \frac{e^{-m_Z|\vec{x} - \vec{y}|}}
         {|\vec{x} - \vec{y}|}
    {\xi_N^c}^\dagger(\vec{y},t)\xi_N(x)
    + h.c.
  \right)~.
  \label{HiggsinoNRL}
\end{eqnarray}
Here,  $\xi_N$ is defined as $\xi_N = (1
-\gamma^0)\tilde{\chi}^0_2e^{imt}/2$, since the sign of the 
the second-lightest neutralino mass is opposite to that of the LSP.

So far we have derived the NR Lagrangian relevant to the neutralino
and chargino scattering. Now we would like to include the terms
relevant to the neutralino pair annihilation into two photons in the
framework. The process $\tilde{\chi}_1^0\tilde{\chi}_1^0\rightarrow
\gamma\gamma$ involves external photons whose momentums are of the
order of the neutralino mass. Those photons cannot be described in the
NR Lagrangian. However, it is possible to calculate the
pair-annihilation cross section in this formalism, by introducing the
a non-unitary four-Fermi terms in the NR Lagrangian. The optical
theorem relates the imaginary part of neutralino forward-scattering
amplitude ${\cal T}$ to the pair-annihilation cross section $\sigma$;
\begin{equation}\label{match}
2~{\rm Im}~{\cal T} = s \sigma v~.
\end{equation}
Here $s$ and $v$ are square of the center-of-mass energy and the
relative velocity, respectively.  The LSP-pair annihilation cross
section therefore can be obtained by calculating the imaginary part of
the forward-scattering amplitude ${\cal T}$ if proper non-unitary
terms are incorporated.  This also justifies the factorization of the
short-distance physics from the the long-distance one, and it is
discussed in detail in Ref.~\cite{Bodwin:1994jh}.

When we explicitly keep the chargino fields in the NR action, the
annihilation into two photons is described by a chargino four-Fermi operator.
The annihilation term is given by
\begin{eqnarray}\label{non-uni}
  {\cal L}_{\rm ann}
  =
  d~
  \eta^\dagger_C(x)
  \xi_C(x)
  ~
  \xi_C^\dagger(x)
  \eta_C(x)~.
\end{eqnarray}
The coupling $d$ is matched to the chargino pair-annihilation cross
section to two photons using Eq.~(\ref{match}), or equivalently
$v\sigma^{(c)}_{\rm ann} = {\rm Im}~d/2$.  We find
$d=i\pi\alpha^2/m^2$. We do not include the annihilation terms for the
LSP or the next-lightest neutralino, since they are responsible to the
processes to $W$- or $Z$-boson pairs, not to photons.

\vspace{1.0cm}


\lromn 3 \hspace{0.2cm} {\bf Two-body State Effective Action}

In order to calculate the neutralino annihilation cross section, it is
convenient to use the effective Lagrangian for the neutralino and the
chargino two-body states than the NR Lagrangian directly. The two-body
state effective action is derived by introducing auxiliary fields for
$\eta_C^\dagger\xi_C$, $\eta_N^\dagger\eta_N^c$,
$\xi_N^\dagger\xi_N^c$ and their Hermitian conjugate fields. After
some calculations, the two-body state effective action is found to be
\begin{eqnarray}
  S_{\rm 2body}
  &=&
  \int \frac{d^4P}{(2\pi)^4}
  \int d^3r
  \left[~~~
    {\phi^{(P)}_C}^\dagger(\vec{r})
    \left(
      E - 2\delta m + \frac{\nabla^2}{m} + \frac{\alpha}{r}
      + \frac{2\pi i \alpha^2}{m^2}\delta{(\vec{r})}
    \right)
    \phi^{(P)}_C(\vec{r})
  \right.
  \nonumber \\
  &&\qquad\qquad\qquad\qquad
  +~\sum_{i = 1}^2
    {\phi^{(P)}_{N_i}}^\dagger(\vec{r})
    \left(E - 2\delta m_i + \frac{\nabla^2}{m}\right)
    \phi^{(P)}_{N_i}(\vec{r})
  \nonumber \\
  &&\qquad\qquad\qquad\qquad
  +~\zeta_C \frac{e^{-m_Zr}}{r}
    {\phi^{(P)}_C}^\dagger(\vec{r})\phi^{(P)}_C(\vec{r})
  \nonumber \\
  &&\qquad\qquad\qquad\qquad
  +~\sum_{i = 1}^2
    \omega_i \frac{e^{-m_Wr}}{r}
    \left\{
    {\phi^{(P)}_C}^\dagger(\vec{r})\phi^{(P)}_{N_i}(\vec{r})
    +
    h.c.
    \right\}
  \nonumber \\
  &&\qquad\qquad\qquad\qquad
  \left.
    +~\zeta_N \frac{e^{-m_Zr}}{r}
      \left\{
        {\phi^{(P)}_{N_1}}^\dagger(\vec{r})\phi^{(P)}_{N_2}(\vec{r})
        +
        h.c.
      \right\}~~~
    \right]~.
  \label{2-body S}
\end{eqnarray}
Here, $\phi^{(P)}_C(\vec{r})$, $\phi^{(P)}_{N_1}(\vec{r})$ and
$\phi^{(P)}_{N_2}(\vec{r})$ correspond to the chargino pair, the
lightest neutralino pair and the second-lightest neutralino pair,
respectively, and $\delta m_1=0$ and $\delta m_2=\delta m_N$.  The
capital $P$ is the center of mass energy and momentum of the two-body
state, and $\vec{r}$ is the relative coordinate. The internal energy
of the neutralino pair, $E$, is defined as $E = P^0 - \vec{P}^2/(4
m)$.  The coupling constants $\omega_1$, $\omega_2$, $\zeta_C$ and
$\zeta_N$ in $S_{\rm 2body}$ are
\begin{equation}
  \omega_1 = -\frac{\sqrt{2}\alpha}{s_W^2}~,
  ~
  \omega_2 = 0~,
  ~
  \zeta_C = -\frac{\alpha c_W^2}{s_W^2}~,
  ~
  \zeta_N = 0~
  \label{parameters}
\end{equation}
for the wino-like  case, and 
\begin{equation} 
  \omega_1 = -\frac{\sqrt{2}\alpha}{4s_W^2}~,
  ~
  \omega_2 = -\frac{\sqrt{2}\alpha}{4s_W^2}~,
  ~
  \zeta_C = -\frac{\alpha(1 - 2c_W^2)^2}{4c_W^2s_W^2}~,
  ~
  \zeta_N = -\frac{\alpha}{4c_W^2s_W^2}~
  \label{parameters1}
\end{equation}
for the Higgsino-like case.

The S-wave states of the neutralino and chargino pairs must give 
the largest contribution to the annihilation cross section,
because higher angular-momentum modes are suppressed by power(s) of
the relative velocity $v$ compared with the S-wave state. 
Thus, the partial wave expansion of  $S_{\rm 2body}$ is convenient for us. 
For this purpose, we expand $\phi^{(P)}_{N_1}$ as
\begin{eqnarray}
  \phi^{(P)}_{N_1}(\vec{r})
  =
  \sum_{l,m}\int\frac{dp}{2\pi} 
  N_{plm}(P)(2p) j_l(pr)Y_{lm}(\theta,\phi)~.
  \label{eigenN}
\end{eqnarray}
Here, $j_l(pr)$ is the spherical Bessel function, and
$Y_{lm}(\theta,\phi)$ is the spherical harmonic function. The quantum
number $p$ is related to the internal energy $E$ as $E = p^2/m$ under
the on-shell condition. In Eq.~(\ref{eigenN}), $N_{plm}$ is the
annihilation operator for the positive-energy state with quantum
numbers $(plm)$.

As mentioned in Section II, the annihilation cross section of the
neutralino pair can be calculated from the forward-scattering
amplitude by using the optical theorem,
\begin{eqnarray}
  \sigma v\left(\frac{p^2}{m}\right)
  =
\frac{1}{2 m^2}~{\rm Im}{\cal T}_{pp}\left(\frac{p^2}{m}\right)
  \label{optical theorem}
\end{eqnarray}
where the forward-scattering amplitude ${\cal T}_{pp}(E)$ is 
\begin{eqnarray}
(2\pi)^4\delta^{(4)}(P-P^\prime)~
i{\cal T}_{pp}(E)  = \frac{8 \pi}{v^2} 
 \vev{N_{p00}(P)N^\dagger_{p00}(P^\prime)}
\end{eqnarray}
and the relative velocity $v = 2p/m$. The amplitude ${\cal T}_{pp}$
may be obtained by solving the equation of motion for the two-body
state derived from Eq.~(\ref{2-body S}), as in the scattering theory in
the quantum mechanics.  Or, we may also calculate it by using the
perturbative theory in a diagrammatic method.

\begin{figure}[t]
  \begin{center}
    \includegraphics[height = 4cm,clip]{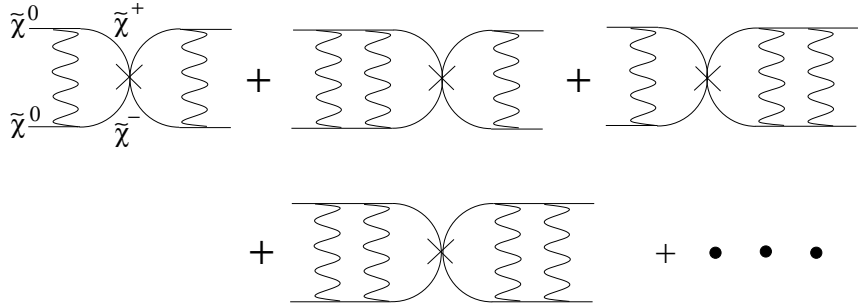}
  \end{center}
  \caption{
Ladder diagrams in the calculation of the forward-scattering amplitude
(${\cal T}_{pp}$). The crossing points correspond to the annihilation
term for chargino.
  \label{ladders}
}
\end{figure}

The dominant contribution to ${\cal T}_{pp}$ in the each loop order
comes from the ladders of photon and weak-boson exchange, as
illustrated in Fig.~\ref{ladders}, provided that the intermediate
neutralinos and charginos are almost on-shell and enhance the
corresponding amplitudes. Similar phenomena are found in QED or QCD as
mentioned in Section I. In the electron and positron pair
annihilation/production at the threshold region, the ratio between the
lowest-order amplitude and the amplitude with additional one-photon
exchange between electron and positron is proportional to $\alpha/v$
at $v\rightarrow 0$. This is well-known as the threshold
singularity. In order to evaluate the  cross section, we
need the resummation of the ladder diagrams since a ladder diagram
with $n$-photon exchange is proportional to $(\alpha/v)^n$. The
other diagrams, such as the crossed ladder diagrams, are suppressed by
additional factors of $v$.

The efficient evaluation for the effect of the resummation of the
ladder diagrams to the electron and positron pair
annihilation/production is to use the wave functions for electron and
positron pair under the QED (Coulomb) potential. When we expand the
two-body state of the electron and positron pair by the wave functions
under the Coulomb potential, the Coulomb potential disappears
from the two-body state action, and the calculation of the
annihilation cross section at the threshold region is only for a
tree-level diagram by the annihilation term, and becomes extremely
simple.

In the evaluation for the neutralino annihilation cross section, the
resummation of weak-boson exchange diagrams, in addition to that
of photon exchange, is required for the heavy neutralino
annihilation cross section. The wave functions of chargino and
neutralino(s) under both Coulomb and Yukawa potentials can be derived
numerically, not analytically. Then, we also consider the behavior of
the cross section in a limit of $m\rightarrow\infty$. In next section
we use the wave functions under the Coulomb potential for the chargino
pair and the free wave functions for the neutralino pairs so that we
can incorporate the all-order ladder diagrams of photon exchange,
and we derive some numerical results.

In order to demonstrate the validity of the effective action
Eq.~(\ref{2-body S}), first, we reproduce the previous one-loop result
for the neutralino annihilation cross section to two photons in our
formulation.  The leading contribution to ${\rm Im}~{\cal T}_{pp}$ in
the perturbative calculation is
\begin{eqnarray}
  {\rm Im}~{\cal T}_{pp}\left(0\right)
  =
  4\pi\alpha^2
  \left|
    (4\pi\omega_1)\int \frac{d^3k}{(2\pi)^3}
    \frac{i}{|\vec{k}|^2 + m_W^2}
    \frac{i}{|\vec{k}|^2/m + 2\delta m}
  \right|^2~.
  \label{1-loop FA}
\end{eqnarray}
The velocity of the incident neutralinos is almost $0$ in
the above calculation since $v/c\sim 10^{-3}$ for the neutralino DM in our
galactic halo. By integrating the r.h.s. of Eq.~(\ref{1-loop FA}) 
and using  Eq.~(\ref{optical theorem}), we find
\begin{eqnarray}
  \sigma v_{\rm 1-loop}(0)
  = 
  \frac{2\pi\alpha^2\omega_1^2}{m_W^2}
  \left(1 + \ds\sqrt{\ds\frac{2m\delta m}{m_W^2}}\right)^{-2}~.
  \label{one-loop result}
\end{eqnarray}
This agrees with the result \cite{Bergstrom:1997fh} which is obtained
of the full one-loop calculation in the heavy wino or Higgsino limit.

It is found from Eq.~(\ref{1-loop FA}) that the amplitude is enhanced
by a factor of $\omega_1 m/m_W$, in which $m$ comes from chargino
propagators in Eq.~(\ref{1-loop FA}) if $\delta m$ can be
neglected. Other factor $1/m_W$ comes from the weak-boson propagators
and the loop integral. The same  enhancement factors also appear in the
higher-order calculations. Whenever one weak-boson exchange is added
to the ladder graph, it gives an additional factor $\omega_1m/m_W$.

Note that the leading correction due to the mass difference $\delta m$
is ${\cal O}(\sqrt{\delta m})$ in Eq.~(\ref{one-loop result}), not
${\cal O}({\delta m})$ as mentioned in the previous section. If the
correction was ${\cal O}({\delta m})$, the integrand in
Eq.~(\ref{1-loop FA}) could be expanded by ${\delta m}$. However, in
this case, the integral is infrared-divergent.

The one-loop cross section is independent of the neutralino mass in
small $\delta m$ limit. On the other hand, the cross section should be
bounded by the unitarity condition $\sigma v < 4\pi/(vm^2)$. Therefore
the higher-order corrections must dominate the cross section, so that
the correct $1/m^2$ behavior would be reproduced. This is explained as
follow. If the neutralino mass is much larger than the weak-boson
mass, we can neglect the effect of the weak-boson mass in the equation
of motion.  This can be seen by introducing the dimensionless
coordinate $x= \alpha_2 m r $ in Eq.~(\ref{2-body S}).  The factor
$e^{-m_Wr}$ can be approximated to 1 in such an extremely large $m$
case.  Then, it is found that the cross section of the extremely heavy
neutralino has the mass dependence $\sigma \sim 1/m^2$ from the
dimensional analysis. In fact, we can solve the equation of motion and
obtain the forward-scattering amplitude analytically in the large mass
limit, because gauge interactions in this Lagrangian generate a common
potential $1/r$, say the Coulomb force.

Now we calculate the cross section in this large mass limit.
We take the wino limit for the LSP as an example, however, the result
is similar for the Higgsino-like neutralino.  It is convenient to
change the basis from ($\phi_C^{(P)}, \phi_N^{(P)}$) to
($\phi_+^{(P)},
\phi_-^{(P)}$) in the two-body state effective action so that the
$1/r$ potential terms are diagonalized as
\begin{eqnarray}
  \left(
    \phi_C^{(P)}~~\phi_N^{(P)}
  \right)
  \left(
    \begin{array}{cc}
      \frac{\alpha + \zeta_C}r & \frac{\omega_1}r
      \\
      \frac{\omega_1}r & 0
    \end{array}
  \right)
  \left(
    \begin{array}{c}
      \phi_C^{(P)}
      \\
      \phi_N^{(P)}
    \end{array}
  \right)
  \rightarrow
  \left(
    \phi_+^{(P)}~~\phi_-^{(P)}
  \right)
  \left(
    \begin{array}{cc}
      \frac{\lambda_+}r & 0
      \\
      0 & \frac{\lambda_-}r
    \end{array}
  \right)
  \left(
    \begin{array}{c}
      \phi_+^{(P)}
      \\
      \phi_-^{(P)}
    \end{array}
  \right)~.
  \label{2-body S of W}
\end{eqnarray}
Here, 
$\lambda_\pm = 0.5\left(\alpha + \zeta_C \pm
\sqrt{(\alpha + \zeta_C)^2 + 4\omega_1^2}\right)$. 
Since $\lambda_+ > 0$ and $\lambda_- < 0$, $\phi_+^{(P)}$ feels an
attractive force, and $\phi_-^{(P)}$ feels a repulsive one. The
neutralino annihilation cross section is derived from the 
$\phi_+^{(P)}$ annihilation, while the contribution from
$\phi_-^{(P)}$ is exponentially suppressed due to  the repulsive
force. Neglecting the $\phi_-$ contribution, the forward-scattering
amplitude is written as follows;
\begin{eqnarray}
  {\rm Im}~{\cal T}_{pp}\left(\frac{p^2}{m}\right)
  \simeq
  \frac{4 \pi^2 \omega_1^2\alpha^2\lambda_+}
  {v (\lambda_+^2 + \omega_1^2)(1 - \exp(-2\pi\lambda_+/v))}~.
\end{eqnarray}
Then, the neutralino annihilation cross section to two photons in the
large mass limit is
\begin{eqnarray}
  \sigma v
  =
  \frac{2\pi^2\omega_1^2\alpha^2\lambda_+}
  {m^2 v (\lambda_+^2 + \omega_1^2)(1 - \exp(-2\pi\lambda_+/v))}
  \sim
  2.8\times 10^{-5}\frac{1}{v m^2}~.
\end{eqnarray}
This cross section behaves as $\sigma v \sim 1/vm^2$, and 
satisfies the unitarity bound as expected. 

\vspace{1cm}


\lromn 4 \hspace{0.2cm} {\bf Higher-order Corrections
in Neutralino Annihilation to Two Photons}

\vspace{0.5cm}
\begin{figure}[t]
  \begin{center}
    \includegraphics[height = 2.4cm,clip]{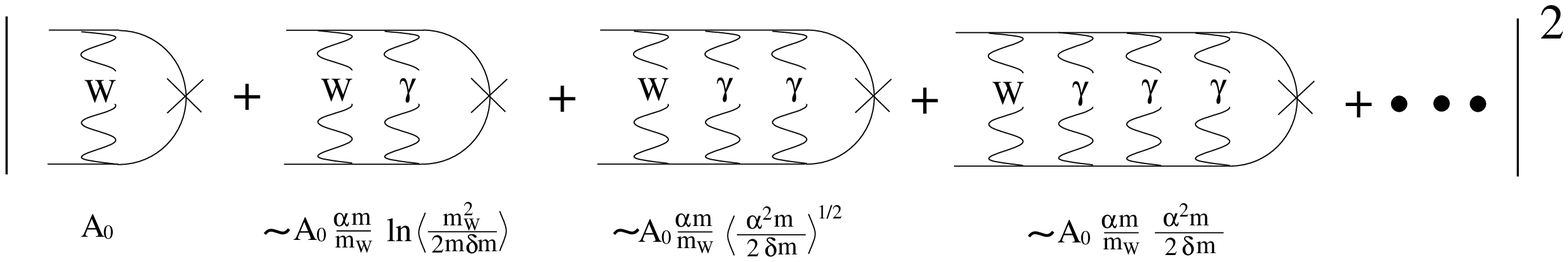}
  \end{center}
  \caption{The ladder diagrams of photon exchange  
relevant to the calculation of the
           forward-scattering amplitude. $A_0$ is the one-loop amplitude;
           $A_0 \sim \alpha\alpha_2 m/m_W$.
  \label{gamma_ladder}
} 
\end{figure}

In the previous section, we showed that the higher-order corrections
from the weak-boson ladder diagrams must be non-negligible if the
neutralino mass $m$ is larger than $m_W/\alpha_2$. In this section we
estimate more carefully the scale where the perturbative approximation
is broken down.  We also discuss another important higher-order
corrections from the photon ladder diagrams.

We start our discussion from the effect of photon exchange. The
corresponding diagrams in the perturbative calculation of ${\cal
T}_{pp}$ are shown in Fig.~\ref{gamma_ladder}. In this figure, the
orders of the amplitudes are also shown. The lowest-order diagram at the
left side of the figure gives the amplitude $A_0$ ($A_0 \sim
\alpha\alpha_2 m/m_W$). The diagrams have an addition factor
$\alpha\sqrt{m/\delta m}$ for each photon exchange, and then, the
higher-order diagrams of photon exchange become important if the mass
difference ($\delta m$) is smaller than $\alpha^2 m$.

The enhancement factor $\alpha\sqrt{m/\delta m}$ comes from the
massless photon exchange at $t$-channel. For the neutralino DM pair
annihilation, the initial neutralinos is almost at rest, and then the
intermediate charginos just slightly deviate from on-shell states by
the mass difference between chargino and neutralino, $\delta m$. The
situation is similar to calculation of the pair production at
threshold in QED or QCD.  In this case, the intermediate particles
deviate from on-shell states by ${\cal O}(v^2 m)$ where  $v$ is the
relative velocity of the incident particles, and the diagrams have an
addition factor $\alpha_{(s)}/v$ for each photon (gluon) exchange.

If neutralino is highly degenerate with chargino in mass, we need
resummation of the photon ladder diagrams. Similar to the positron and
electron annihilation/production explained in the previous section,
the resummation of the ladder diagrams is possible analytically, by
expanding the chargino-pair field ($\phi_{C}(\vec{r})$) in the
two-body state effective action Eq.~(\ref{2-body S}), in terms of the
eigenfunctions of the Schr\"{o}dinger equation with the Coulomb
potential,
\begin{eqnarray}
  \phi^{(P)}_C(\vec{r})
  &=&
  \sum_{l,m}
  \left[
    \int\frac{dk}{2\pi} C_{klm}(P)R_C^{(kl)}(r)
    +
    \sum_n \tilde{C}_{nlm}(P)\tilde{R}_C^{(nl)}(r)
  \right]Y_{lm}(\theta,\phi)~.
  \label{eigen fn}
\end{eqnarray}
Here, $C_{klm}$ is the annihilation  operator of the continuum state of a
chargino pair with quantum numbers $(klm)$, and $\tilde{C}_{nlm}$ is
one of a chargino-pair bound state with $(nlm)$. $R_C^{(kl)}(r)$ and
$\tilde{R}_C^{(nl)}(r)$ are eigenfunctions of the radial direction for
the states. Both $R_C$ and $\tilde{R}_C$ are given by the confluent
hypergeometric function, and the S-wave parts of these functions
are  \cite{randau}
\begin{eqnarray}
  R_C^{(k0)}(\vec{r})
  &=&
  2\sqrt{\frac{\pi\alpha m k}{1 - \exp(-\pi \alpha m/k)}}
  e^{-ikr}
  ~_1F_1\left(\frac{\alpha m}{2k}i + 1, 2, 2ikr\right)~,
  \nonumber \\
  \tilde{R}_C^{(n0)}(\vec{r})
  &=&
  2\left(\frac{\alpha m}{2n}\right)^{3/2}
  e^{-\alpha m r/2n}
  ~_1F_1\left(- n + 1, 2, \frac{\alpha m r}{n}\right)~,
  \label{S eigen fn}
\end{eqnarray}
where $~_1F_1$ is the (Kummer's) confluent hypergeometric function.
With this expansion and Eq.~(\ref{eigenN}), we derive the S-wave
effective Lagrangian of the two-body states.  We show details of the
Lagrangian in appendix A.

In the following we calculate the higher-order corrections of 
gauge-boson exchange.  The forward-scattering amplitude ${\cal
T}_{pp}$ is calculated from the S-wave effective Lagrangian by
treating weak-boson exchange interactions perturbatively. The
effect of all-order photon-exchange effects are now automatically
taken into account in $\alpha$ dependent coefficients of the
interactions.

We first calculate the cross section in the leading level where the
effect of only one $W$-boson exchange is included. After some
calculation, we find
\begin{eqnarray}
  \sigma v\left(\frac{p^2}{m}\right)
  &=&
  \frac{2\pi\alpha^2}{m^2}
  \left|
    \frac{i\omega_1}{p}\int\frac{dk}{2\pi}
    \frac{im A_{NC}(p,k)}{p^2 - k^2 -2m\delta m + i0^+}
    \sqrt{\frac{\pi\alpha m k}{1 - \exp(-\pi \alpha m/k)}}
  \right.
  \nonumber \\
  &~&
  \qquad\qquad +
  \left.
    \frac{i\omega_1}{p}\sum_n
    \frac{im A_{N\tilde{C}}(p,n)}
    {p^2 + \alpha^2m^2/4n^2 - 2m\delta m + i0^+}
    \left(\frac{\alpha m}{2}\right)^{3/2}~~
  \right|^2~.
  \nonumber
\end{eqnarray}
The coefficients $A_{NC}$ and $A_{N\tilde{C}}$ are defined in Appendix
A. Here the first term in r.h.s. comes from the continuum states of
chargino pair, while the second term from the discrete states of
chargino pair.

We first investigate contributions of the chargino bound states to the
cross section. We can see from Eq.(\ref{cs}) that their contributions
are suppressed by a factor $\alpha$ compared with the continuum part.
This is because the overlap between the wave functions of the
neutralino pair and the chargino bound states is suppressed by
$\alpha$.  If energy of the neutralino pair is on the pole of the
chargino bound state ($p^2/m \sim 2\delta m -\alpha^2m/4n^2$), the
cross section is enhanced beyond the suppression due to the small
overlap of the wave functions. However, it is very unlikely for the
pair annihilation of the neutralino DM. The average DM velocity in our
galactic halo is roughly $v/c \sim 10^{-3}$. Thus, the bound state
contribution is negligible as far as the neutralino mass is smaller than
100TeV, since $p^2/m\sim 10^{-6}m$ and the mass difference $\delta m$
may be larger than 100MeV.

Thus, we only have to take into account the contribution from the
continuum chargino-pair states. Since the continuum part of the cross
section is a smooth function around the $p \sim 0$, the cross section
for the neutralino DM annihilation may be approximated by $\sigma v
(0)$. The result is
\begin{eqnarray}
  \sigma v (0)
  &=&
  \frac{2\pi\alpha^2}{m^2}\left|C(0)\right|^2
  \label{cs0}
\end{eqnarray}
where
\begin{eqnarray}
  C(0)
  &=&
  i\omega_1\int\frac{dk}{2\pi}
  \frac{4im~\pi\alpha m k}{1 - e^{-\pi \alpha m/k}}
  \ds\frac{\exp
    \left[
      -\ds\frac{\alpha m}{2k}\arctan(\frac{2m_Wk}{m_W^2 - k^2})
    \right]}
    {k^2 + m_W^2}
  \frac{1}{- k^2 -2m\delta m}~.
  \nonumber
\end{eqnarray}
In a limit of vanishing $\alpha$, the $\sigma v$ reduces to $\sigma
v_{\rm 1-loop}$ in Eq.~(\ref{one-loop result}).

\begin{figure}[t]
 \includegraphics[height = 4.6cm,clip]{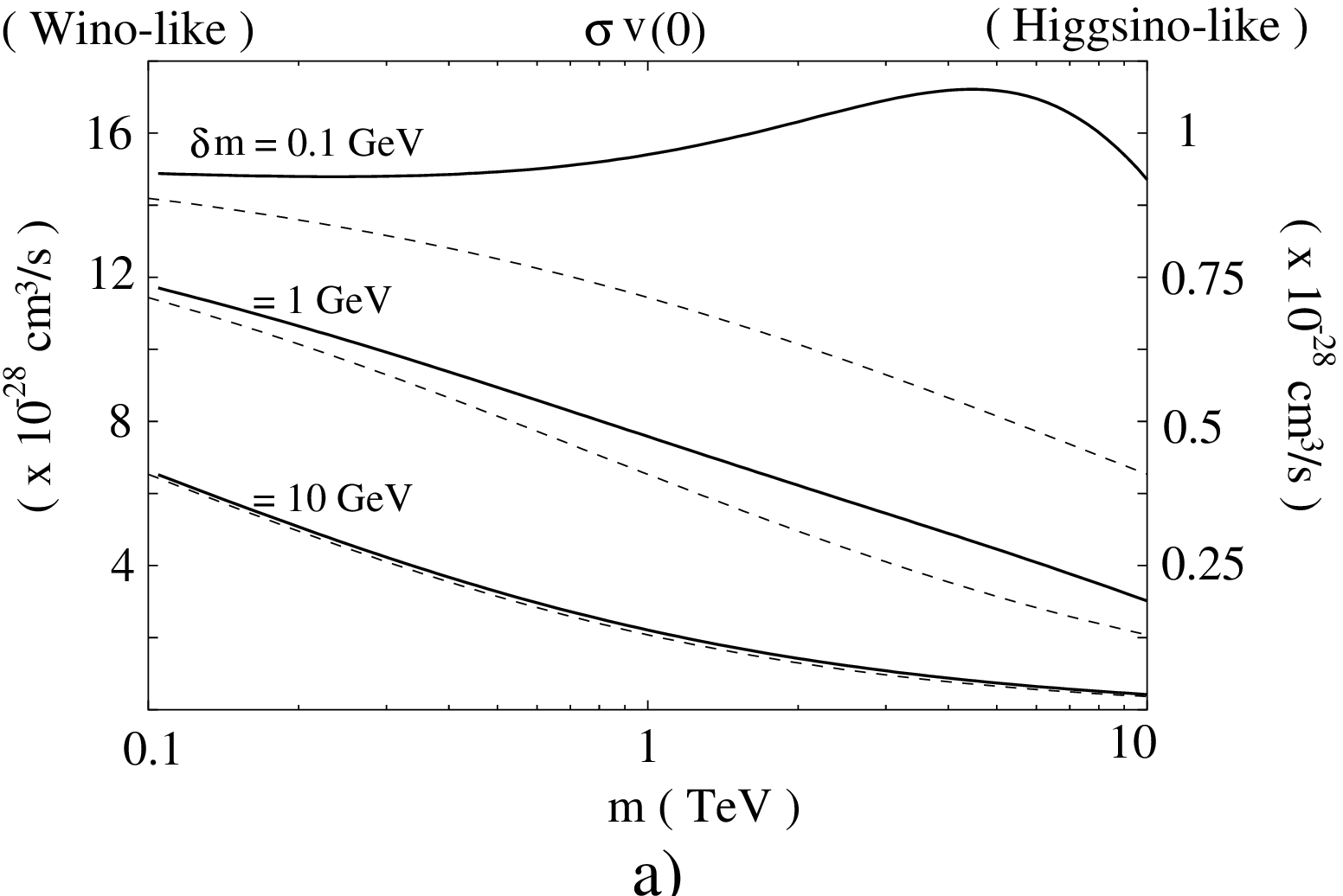} ~~~
 \includegraphics[height = 5.0cm,clip]{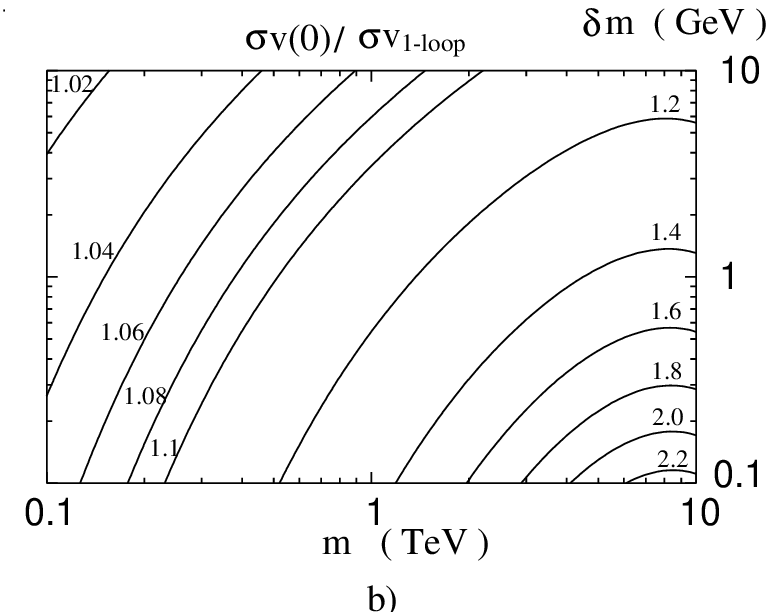} 
\caption{ 
 \label{leading} a) The leading
 $\tilde{\chi}^0_1\tilde{\chi}^0_1\rightarrow 2\gamma$ cross section
 taking into account all-order QED effect, $\sigma v$, as a function
 of the neutralino mass $m$ (solid lines). For comparison, result of
 the one-loop calculation Eq.~($\ref{one-loop result}$) is also shown
 (dashed lines). b) The contours of  $\sigma/\sigma_{\rm
 1-loop}$ in a ($m$, $\delta m$) plane.
}
\end{figure}

In Fig.~\ref{leading} a), solid lines are $\sigma v(0)$ for $\delta
m=0.1$GeV, 1 GeV and 10 GeV. The left axis corresponds to the
wino-like neutralino annihilation cross section, and the right axis is
for the Higgsino-like one. For comparison, $\sigma v_{\rm 1-loop}$ in
Eq.~($\ref{one-loop result}$) are also shown by dashed lines. Note
that influences of the QED potential always work to enhance the cross
section. This is because the Coulomb force acts as an attractive force
between the chargino pair. In Fig.~\ref{leading} b), contours of
$\sigma v(0)/\sigma v_{\rm 1-loop}$ is shown in a ($m$, $\delta m$)
plane. The enhancement is large if the mass difference $\delta m$ is
small and $m$ is large as expected. For the wino-like neutralino,
which is highly degenerate with chargino as $\delta m\sim {\cal
O}(100)$MeV, the enhancement by factor of 2 is possible when the
neutralino mass is $ {\cal O}(1)$TeV.

\begin{figure}[t]
 \includegraphics[height = 5cm,clip]{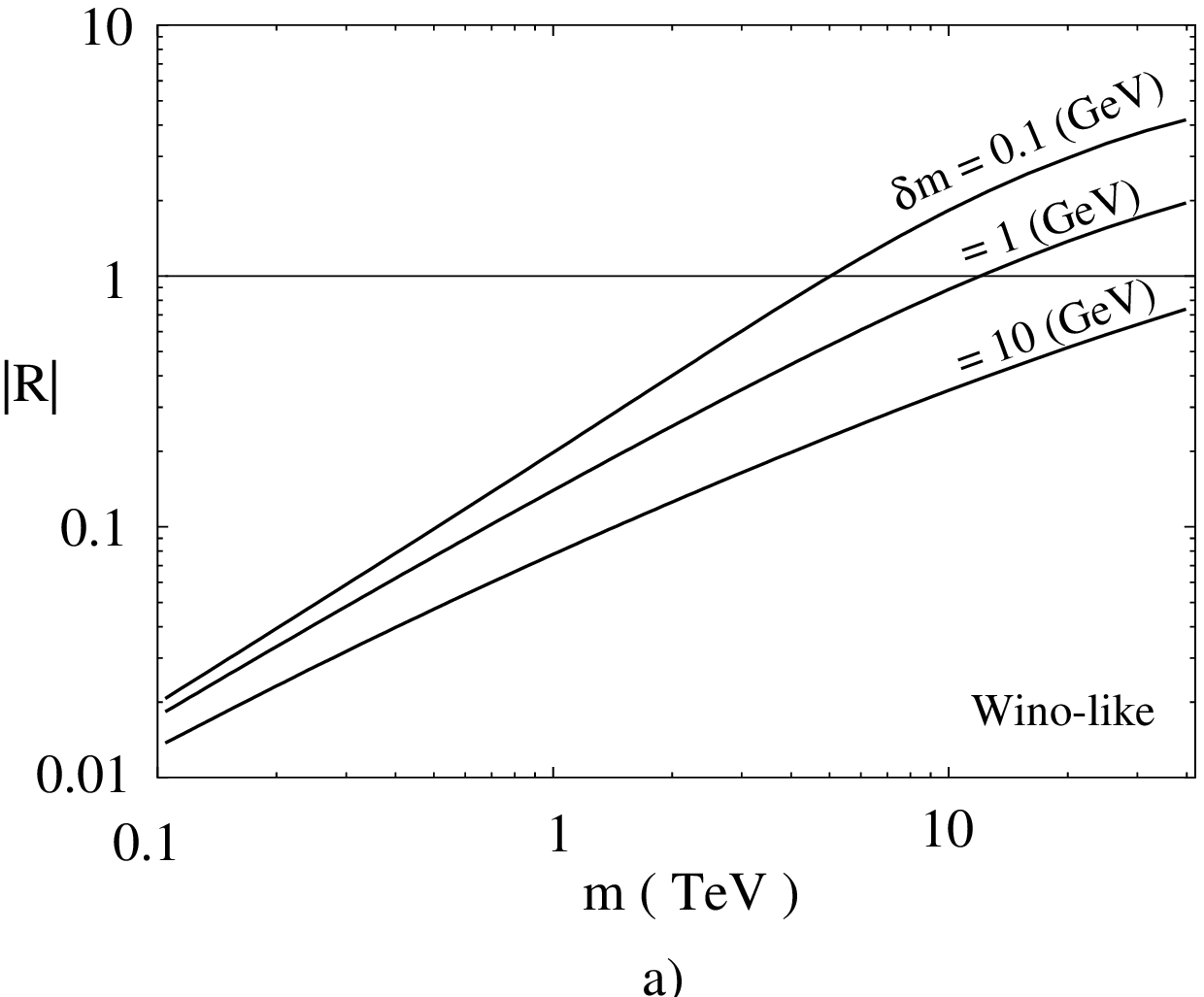}
 \qquad\qquad
 \includegraphics[height = 5cm,clip]{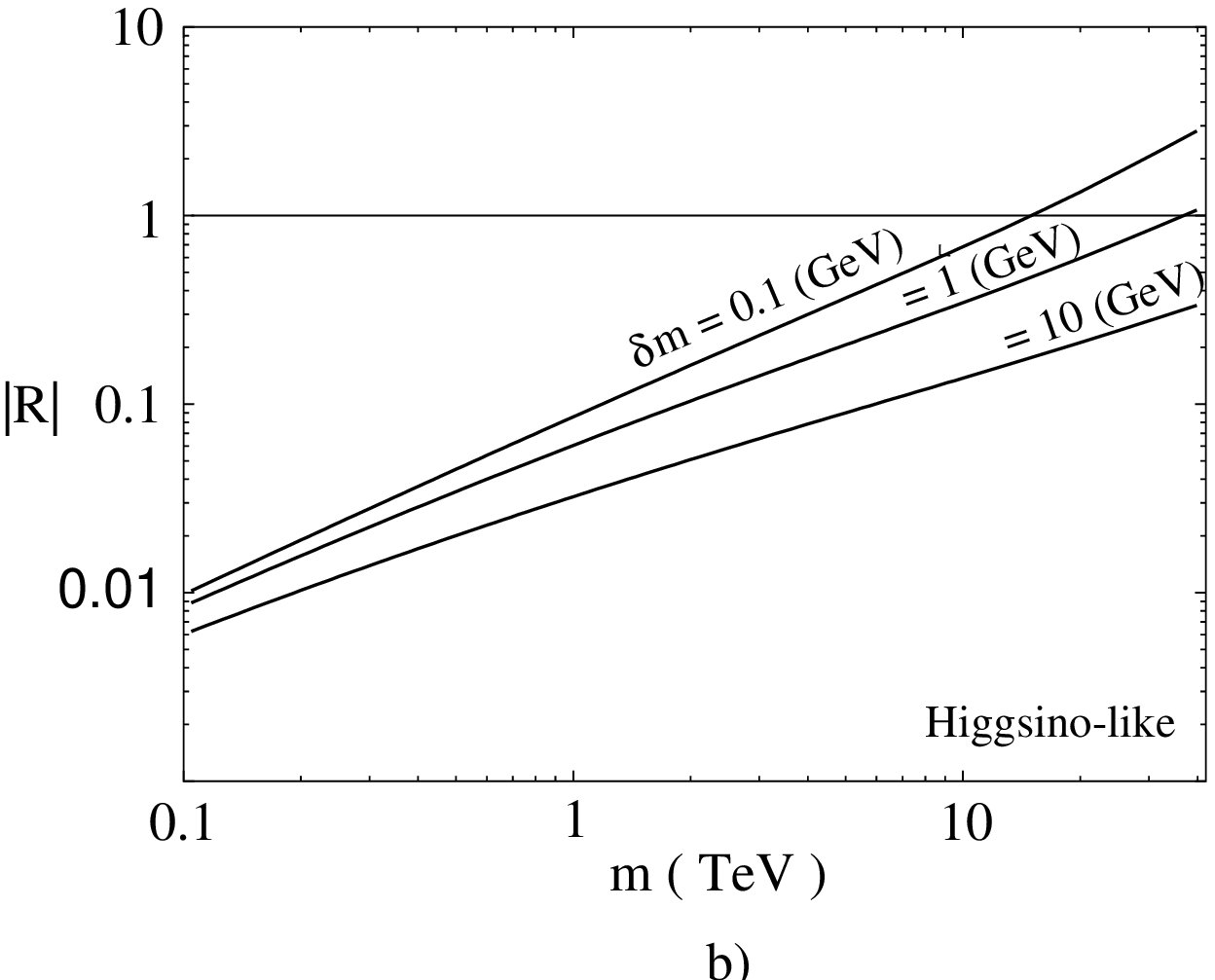}
 \caption{
 The ratio between the next-to-leading order amplitude and the
 leading-order amplitude as a function of the neutralino mass $m$. a)
 and b) are for the wino- and  the Higgsino-like
 neutralino, respectively.
  \label{next leading} }
\end{figure}

\begin{figure}[t]
  \begin{center}
    \includegraphics[height = 2.8cm,clip]{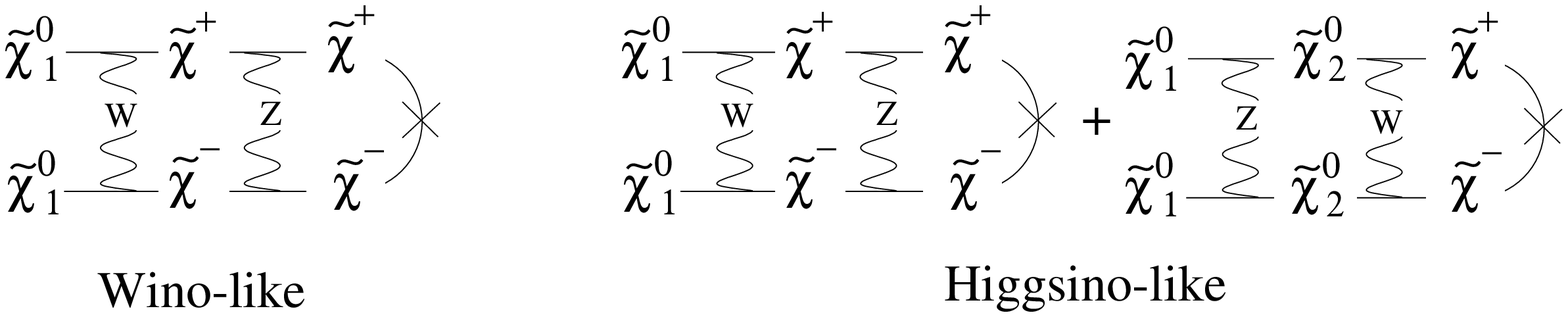}
  \end{center}
   \caption{
Diagrams which contribute to next-to-leading order correction. a)
and b) are for the wino- and  Higgsino-like
neutralino DM, respectively.
 \label{2-loop diagrams} }
\end{figure}

Next, we discuss the  next-to-leading order corrections arising 
from the diagrams of two weak-boson exchange. The detail of the 
calculation will be given elsewhere \cite{work_in_progress}, and 
we only show the ratio between  the next-to-leading order corrections 
and the  leading amplitude, $R$, in Fig.\ref{next leading}. 

In the wino-like case, the $Z$ exchange between the chargino pair,
shown in Fig.~\ref{2-loop diagrams}, leads to the next-to-leading
order corrections. In Fig.~\ref{next leading} a), the ratio $R$ in the
wino-like case is plotted as a function of $m$ with $\delta m$ = 0.1
GeV and 1 GeV. We call the value of $m$ where $|R|=1$ as the critical
neutralino mass $m_{\rm crit}$.  Above and around the mass, the
higher-order corrections is larger than the leading-order one, therefore
the perturbation is not reliable any more. For the wino-like case,
$m_{\rm crit}\sim 10$TeV, and for the Higgsino-like case
$m_{\rm crit}\sim$ $O(10)$TeV.

\vspace{1cm}


\lromn 5 \hspace{0.2cm} {\bf Conclusion and Discussion}

\vspace{0.5cm}

Although the existence of DM is established, we do not yet know the DM
nature.  The neutralino DM predicted in MSSM is actively searched for
by various experiments, trying to solve this myth.  The search for
gamma-ray signal from the center of our galaxy is a feasible way to
discover the neutralino DM. Especially the monochromatic gamma ray
from the neutralino DM pair annihilation is a robust signal.

The reliable estimation of the signal rate is important to either set
limit of the neutralino DM or interpret the observed signals.  The
major systematic error of the signal rate comes from the distribution
of the dark matter in our galaxy.  However, the pair annihilation
cross section $\sigma(\chi_1^0\chi_1^0\rightarrow 2\gamma)$ has
not been also  understood completely. The problem arises when
$\tilde{\chi}^0_1$ is wino- or Higgsino-like. In that case, the
one-loop cross section $\sigma v_{\rm 1-loop}$ is approximately
constant, and does not scale as $1/m^2$.  The one-loop cross section
obviously breaks the unitarity bound in $m\rightarrow \infty$
limit. The enhancement comes from chargino in the loop, which is
degenerate with the LSP neutralino.

In this paper we study the higher-order loop corrections to the pair
annihilation process. We find that the dominant contribution comes
from the ladder diagram of weak bosons and photon exchange. We study the
corrections using NRQED technique, which allows us to incorporate
ladder-type QED corrections analytically to all orders, while including
weak boson exchange corrections perturbatively.  We find the QED
corrections enhance the cross section up to a factor of two.  We also
find the critical scale $m_{\rm crit}$, around and above which the
next-to-leading order corrections by weak-boson exchange are larger
than the leading contribution, is $\sim$  $O(10)$ TeV for the Higgsino-like
LSP and $\sim$ 10 TeV for the wino-like LSP. The corrections reduce
the cross section for $m< m_{\rm crit}$. We also find the
pair-annihilation cross section satisfies the unitarity bound in a
limit of $m\rightarrow
\infty$. We note that our formulation may be used for any other
DM candidate if the DM comes from SU(2) multiplets and the DM is
degenerate with other components in mass.

The relic density of the Higgsino- or wino-like LSP with the mass above
1 TeV may be consistent to the matter density of the Universe
$\Omega_M$. The monochromatic gamma ray in TeV region may be searched
for the future atmospheric Cherenkov telescopes (ACT)
\cite{Bergstrom:1997fj}. The signal flux may be above the sensitivity,
if the DM density distribution is singular at the center of our
galaxy.  Therefore it is important to estimate the neutralino
pair-annihilation cross section reliably. For such a heavy Higgsino- or
wino-like neutralino, one has to sum the weak corrections  to all
orders as we discussed in this paper. The resummation is possible by
the solving the NR equation of motion of neutralino \cite{work_in_progress}.

\underline{Acknowledgments}

We thank Dr. Gondolo, Dr. Kinoshita, Dr.~Onogi and Dr.~Sumino for useful
discussion, and Dr.~Fukae for giving his master thesis to us.  The
work of SM is supported by JSPS.  This work is supported in part by
the Grant-in-Aid for Science Research, Ministry of Education, Science
and Culture, Japan (No.~14046225 and No.~1313527 for JH, and
No.~14540260 and No.~14046210 for MMN).

\vspace{0.7cm}


{\bf Appendix A S-wave Lagrangian}

\vspace{0.5cm}

In this appendix, we list the full S-wave Lagrangian, which is used to
calculate the leading and the next-to-leading order corrections of the
cross section.  The S-wave Lagrangian is obtained by substituting
Eq.~(\ref{eigenN}) and Eq.~(\ref{eigen fn}) to the two-body state
effective action Eq.~(\ref{2-body S}), and extracting the S-wave
parts. We find 
\begin{eqnarray}
  {\cal L}_S
  &=&
  {\cal L}_0 + {\cal L}_W + {\cal L}_Z + {\cal L}_{\rm a}~,
  \nonumber 
\end{eqnarray}
where 
\begin{eqnarray}
  {\cal L}_0
  &=&
  \int\frac{dp}{2\pi}
  \left(E - \frac{p^2}{m}\right)
  {N^{(1)}_p}^\dagger N^{(1)}_p
  +
  \int\frac{dp}{2\pi}
  \left(E - 2\delta m_N - \frac{p^2}{m}\right)
  {N^{(2)}_p}^\dagger N^{(2)}_p
  \nonumber \\
  &+&
  \int\frac{dk}{2\pi}
  \left(E - 2\delta m - \frac{k^2}{m}\right)
  C^\dagger_k C_k
  +
  \sum_n 
  \left(E - 2\delta m + \frac{m\alpha^2}{4n^2}\right)
  \tilde{C}^\dagger_n \tilde{C}_n~,
  \nonumber \\
  \nonumber \\
  {\cal L}_W
  &=&
  \sum_{i = 1}^2
  \omega_i \int \frac{dp}{2\pi}\frac{dk}{2\pi}
  A_{NC}(p,k)
  \left[
    C^\dagger_k N^{(i)}_p + {N^{(i)}_p}^\dagger C_k
  \right]
  \nonumber \\
  &+&
  \sum_{i = 1}^2
  \omega_i \sum_n \int \frac{dp}{2\pi}
  A_{N\tilde{C}}(p,n)
  \left[
    \tilde{C}^\dagger_n N^{(i)}_p + {N^{(i)}_p}^\dagger \tilde{C}_n
  \right]~,
  \nonumber \\
  \nonumber \\
  {\cal L}_Z
  &=&
  \zeta_C \int \frac{dk}{2\pi}\frac{dk'}{2\pi}
  A_{CC}(k,k')
  C^\dagger_k C_{k'}
  +
  \zeta_C \int \sum_{n,n'}
  A_{\tilde{C}\tilde{C}}(n,n')
  \tilde{C}^\dagger_n \tilde{C}_{n'}
  \nonumber \\
  &+&
  \zeta_C \sum_n \int \frac{dk}{2\pi}
  A_{C\tilde{C}}(k,n)
  \left[
    \tilde{C}^\dagger_n C_k + C_k^\dagger \tilde{C}_n
  \right]
  \nonumber \\
  &+&
  \zeta_N \int \frac{dp}{2\pi}\frac{dp'}{2\pi}
  A_{NN}(p,p')
  \left[
    {N^{(1)}_p}^\dagger N^{(2)}_{p'} + {N^{(2)}_{p'}}^\dagger N^{(1)}_p
  \right]~,
  \nonumber \\
  \nonumber \\
  {\cal L}_a
  &=&
  i\frac{2\alpha^2}{m^2}
  \left[
    \int \frac{dk}{2\pi}~
    \sqrt{\frac{\pi\alpha m k}{1 - \exp(-\pi \alpha m/k)}}
    C_k^\dagger~
    +
    \sum_n \left(\frac{\alpha m}{2 n}\right)^{3/2}
    \tilde{C}_n^\dagger
  \right]
  \nonumber \\
  \nonumber \\
  &~& ~~~ \cdot
  \left[
    \int \frac{dk'}{2\pi}
    \sqrt{\frac{\pi\alpha m k'}{1 - \exp(-\pi \alpha m/k')}}
    C_{k'}
    +
    \sum_{n'} \left(\frac{\alpha m}{2 n'}\right)^{3/2}
    \tilde{C}_{n'}
  \right]~.
  \label{S L}
\end{eqnarray}
Here, we omit the arguments $P$, $l$, and $m$ for simplicity, which
are the center of mass energy and momentum, and orbit-angular
momentums, respectively. Coupling constants $\omega_1$, {\it etc}, are
defined by the Eqs.~(\ref{parameters},\ref{parameters1}). $N^{(1)}_p$
and $N^{(2)}_p$ are annihilation operators of the lightest and
next-lightest neutralino pairs, respectively, and $C_k$ and $\tilde{C}_n$
are for the continuum and discrete chargino pair states.  The coefficients
$A_{NC}$ and $\tilde{A}_{N\tilde{C}}$, {\it etc}, are given by 
a little complicated functions,
\begin{eqnarray}
  A_{NC}(p,k)
  &=&
  -\frac{4}{\alpha m}\sqrt{\frac{\pi \alpha m k}{1 - \exp(-\pi \alpha m/k)}}
  {\rm Im}
  \left[
    \left(
      \frac{m_W - i(p + k)}{m_W - i(p - k)}
    \right)^{-i\alpha m/2k}
  \right]~,
  \nonumber \\
  \nonumber \\
  A_{N\tilde{C}}(p,n)
  &=&
  -\frac{4}{\alpha m}\left(\frac{\alpha m}{2n}\right)^{3/2}
  {\rm Im}
  \left[
    \left(
      \frac{m_W - \alpha m/2n - ip}{m_W + \alpha m/2n - ip}
    \right)^n
  \right]~,
  \nonumber \\
  \nonumber \\
  A_{CC}(k,k')
  &=&
  \frac{4}{m_z^2 + (k - k')^2}
  \sqrt{\frac{\pi \alpha m k}{1 - \exp(-\pi \alpha m/k)}}
  \sqrt{\frac{\pi \alpha m k'}{1 - \exp(-\pi \alpha m/k')}}
  \nonumber \\
  \nonumber \\
  &\times&
  \left(
    \frac{m_z + i(k + k')}{m_z - i(k - k')}
  \right)^{i\alpha m/2k}
  \left(
    \frac{m_z + i(k + k')}{m_z - i(k' - k)}
  \right)^{i\alpha m/2k'}
  \nonumber \\
  \nonumber \\
  &\times&
  ~_2F_1
  \left(
    \frac{\alpha m}{2k}i + 1,\frac{\alpha m}{2k'}i + 1, 2,
    -\frac{4kk'}{m_z^2 + (k - k')^2}
  \right) ~,
  \nonumber \\
  \nonumber \\
  A_{\tilde{C}\tilde{C}}(n,n')
  &=&
  \frac{2}{4m_z^2 - \alpha^2m^2(1/n - 1/n')^2}
  \frac{\alpha^3m^3}{(nn')^{3/2}}
  \nonumber \\
  \nonumber \\
  &\times&
  \left(
    \frac{2m_z - \alpha m(1/n - 1/n')}{2m_z + \alpha m(1/n + 1/n')}
  \right)^n
  \left(
    \frac{2m_z - \alpha m(1/n' - 1/n)}{2m_z + \alpha m(1/n + 1/n')}
  \right)^{n'}
  \nonumber \\
  \nonumber \\
  &\times&
  ~_2F_1
  \left(
    1 - n, 1 - n', 2,
    \frac{4\alpha^2m^2/(nn')}{4m_z^2 + \alpha^2m^2(1/n - 1/n')^2}
  \right)~,
  \nonumber \\
  \nonumber \\
  A_{C\tilde{C}}(k,n)
  &=&
  \frac{4}{m_z^2 + (ik - \alpha m/(2n))^2}
  \sqrt{\frac{\pi \alpha m k}{1 - \exp(-\pi \alpha m/k)}}
  \left(\frac{\alpha m}{2 n}\right)^{3/2}
  \nonumber \\
  \nonumber \\
  &\times&
  \left(
    \frac{m_z + \alpha m/(2n) + ik}{m_z + \alpha m/(2n) - ik}
  \right)^{i\alpha m/2k}
  \left(
    \frac{m_z + \alpha m/(2n) + ik}{m_z - \alpha m/(2n) + ik}
  \right)^{-n}
  \nonumber \\
  \nonumber \\
  &\times&
  ~_2F_1
  \left(
    \frac{\alpha m}{2k}i + 1,1 - n, 2,
    \frac{2ik\alpha m/n}{m_z^2 - (\alpha m/(2n) - ik)^2}
  \right)~,
  \nonumber \\
  \nonumber \\
  A_{NN}(p,p')
  &=&
  \log
  \left(
    \frac{m_z^2 + (p + p')^2}{m_z^2 + (p - p')^2}
  \right)~.
  \label{cs}
\end{eqnarray}
Here, $~_2F_1$ is the (Gauss's) hypergeometric function. 

\vspace{1cm}


\end{document}